\documentclass[a4paper,aps,prd,10pt,preprintnumbers,showpacs,twocolumn,superscriptaddress,nofootinbib,amsmath,amssymb]{revtex4-1}
\usepackage{graphicx}
\usepackage{cmap}
\usepackage[utf8]{inputenc}
\usepackage[T1]{fontenc}

\def\imo{i}

\def\K{{\cal K}}

\begin{document}
\title{Ringing of the Extreme Regular Black Holes}
\author{Milena Skvortsova}\email{milenas577@mail.ru}
\affiliation{Peoples' Friendship University of Russia (RUDN University), 6 Miklukho-Maklaya Street, Moscow, 117198, Russia}
\begin{abstract}
We investigate quasinormal ringing in both time and frequency domains for scalar and neutrino perturbations around black hole solutions that simultaneously describe regular and extreme configurations within a non-linear electrodynamics framework. Two types of solutions are considered: those with de Sitter and Minkowski cores. The quasinormal frequencies obtained from two independent methods exhibit excellent agreement. Furthermore, we derive an analytical expression in the eikonal limit and discuss the validity of the correspondence between eikonal quasinormal modes and null geodesics.
\end{abstract}
\maketitle
\section{Introduction}

Quasinormal modes (QNMs) of black holes are characteristic oscillations that dominate the response of a black hole to perturbations \cite{Konoplya:2011qq,Kokkotas:1999bd}. These modes are solutions to the perturbed Einstein equations that satisfy specific boundary conditions: purely ingoing waves at the event horizon and purely outgoing waves at infinity. They represent damped oscillations due to the emission of gravitational waves, with frequencies and damping times determined by the black hole's mass, charge, and angular momentum. Studying QNMs provides crucial insights into the stability of black holes, the nature of spacetime in strong gravitational fields, and tests of general relativity and alternative theories of gravity. QNMs are also of significant interest in the context of gravitational wave astronomy, as they are expected to be observable in the signals detected by observatories like LIGO and Virgo \cite{LIGOScientific:2016aoc,LIGOScientific:2020zkf}.

Special attention has been paid to the quasinormal spectra of extreme and near-extreme black holes, where, for example, the electric charge or angular momentum approaches their maximal values. The spectrum of the extremally charged Reissner-Nordström black hole is remarkable because various types of perturbations can be obtained from each other via supersymmetric transformations that preserve the extreme black hole \cite{Onozawa:1995vu,Onozawa:1996ba}. Another interesting aspect is that extremality brings potential instability to the exactly extreme state, a topic discussed in numerous works \cite{Aretakis:2013dpa,Aretakis:2011ha,Durkee:2010ea,Aretakis:2012ei,Aretakis:2011hc,Aretakis:2011gz,Lucietti:2012sf}. Instability may also occur for highly charged near-extreme black holes \cite{Konoplya:2013sba}.

Another intriguing property of some black hole models is the possibility of avoiding the central singularity, which indicates the limits of General Relativity. The quasinormal modes of regular black holes have been extensively studied in recent years \cite{Bronnikov:2019sbx,Konoplya:2024hfg,Bolokhov:2023ruj,Pedrotti:2024znu,Li:2016oif,Li:2022kch,Zhang:2024nny,Huang:2015cha,Lin:2013ofa,Panotopoulos:2019qjk,Rayimbaev:2022mrk,Wahlang:2017zvk,Myrzakulov:2023rkr,Jha:2023wzo,Fernando:2012yw,Meng:2022oxg,Guo:2024jhg,Dubinsky:2024aeu}. It has been shown that regular black holes may exhibit distinctive features in their spectra, such as purely imaginary quasinormal modes or outbursts of overtones \cite{Konoplya:2022hll}. Recently, it was demonstrated that a broad class of regular black holes can be obtained via an infinite series of higher curvature corrections \cite{Bueno:2024dgm}, suggesting that the solution to the singularity problem may arise from string theory corrections to classical gravity.

Here, we consider black hole models that incorporate both extremality and regularity, as recently obtained in \cite{Bronnikov:2024izh}. Extreme charged black holes in the Einstein-Maxwell theory are typically singular; however, regularity was achieved in \cite{Bronnikov:2024izh} through non-linear modifications of Maxwell electrodynamics. We will study the quasinormal spectra of two models of regular and extreme black holes: those with Minkowski and de Sitter cores.

The paper is organized as follows. In Sec. \ref{sec:wavelike}, we summarize the main information about the metric, wave-like equations, and effective potentials. Sections \ref{sec:WKB} and \ref{sec:TD} provide brief reviews of the methods used for calculating quasinormal modes. Sections \ref{sec:QNM} and \ref{sec:Eikonal} discuss the calculations of quasinormal modes, including analytical expressions in the eikonal regime. In Sec. \ref{sec:conclusions}, we summarize the obtained results and mention some remaining problems.

\section{Extreme regular black holes}\label{sec:wavelike}


The metric of the Bronnikov extremally charged black holes within non-linear electrodynamics is given by the following line element \cite{Bronnikov:2024izh}:
\begin{equation}\label{metric}
  ds^2=-f(r)dt^2+\frac{dr^2}{f(r)}+r^2(d\theta^2+\sin^2\theta d\phi^2),
\end{equation}
where the metric function for the regular black hole with a Minkowski core is
$$
\begin{array}{rcl}
f_{1}(r)&=&\displaystyle \left(1-\frac{M r^3}{\left(a ^2+r^2\right)^2}\right)^2,\\
\end{array}
$$
and for the black hole with a de Sitter core, it is
$$
\begin{array}{rcl}
f_{2}(r)&=&\displaystyle \left(1-\frac{M r^2}{\left(a ^2+r^2\right)^{3/2}}\right)^2.\\
\end{array}
$$
Here, $a$ is the parameter characterizing the non-linearity, and $M$ represents the ADM mass. Throughout this work, we will measure all dimensional quantities in units of mass, setting $M=1$. The event horizon exists for 
$$a \lessapprox 0.32,$$ in the model with a Minkowski core and for $$a \lessapprox 0.38$$ in the model with a de Sitter core.

The general relativistic equations for the scalar ($\Phi$) and Dirac ($\Upsilon$) fields can be expressed as follows:
\begin{subequations}\label{coveqs}
\begin{eqnarray}\label{KGg}
\frac{1}{\sqrt{-g}}\partial_\mu \left(\sqrt{-g}g^{\mu \nu}\partial_\nu\Phi\right)&=&0, 
\\\label{covdirac}
\gamma^{\alpha} \left( \frac{\partial}{\partial x^{\alpha}} - \Gamma_{\alpha} \right) \Upsilon&=&0,
\end{eqnarray}
\end{subequations}
where $\gamma^{\alpha}$ are the non-commutative gamma matrices and $\Gamma_{\alpha}$ represent the spin connections in the tetrad formalism.
Upon separation of variables in the background metric (\ref{metric}), the equations in (\ref{coveqs}) reduce to the Schrödinger-like wave equation \cite{Kokkotas:1999bd,Berti:2009kk,Konoplya:2011qq}:
\begin{equation}\label{wave-equation}
\dfrac{d^2 \Psi}{dr_*^2}+(\omega^2-V(r))\Psi=0,
\end{equation}
where the "tortoise coordinate" $r_*$ is defined as:
\begin{equation}\label{tortoise}
dr_*\equiv\frac{dr}{f(r)}.
\end{equation}

The effective potential for the scalar field is given by
\begin{equation}\label{potentialScalar}
V(r)=f(r)\frac{\ell(\ell+1)}{r^2}+\frac{1}{r}\cdot\frac{d^2 r}{dr_*^2},
\end{equation}
where $\ell=0, 1, 2, \ldots$ are the multipole numbers.
For the Dirac field ($s=1/2$), there are two isospectral potentials:
\begin{equation}
V_{\pm}(r) = W^2\pm\frac{dW}{dr_*}, \quad W\equiv \left(\ell+\frac{1}{2}\right)\frac{\sqrt{f(r)}}{r}.
\end{equation}
The isospectral wave functions can be transformed into one another by the Darboux transformation:
\begin{equation}\label{psi}
\Psi_{+}\propto \left(W+\dfrac{d}{dr_*}\right) \Psi_{-},
\end{equation}
indicating that it is sufficient to calculate quasinormal modes for only one of the effective potentials. We will focus on $V_{+}(r)$ because the WKB method performs better with this potential.

\begin{figure}
\resizebox{\linewidth}{!}{\includegraphics{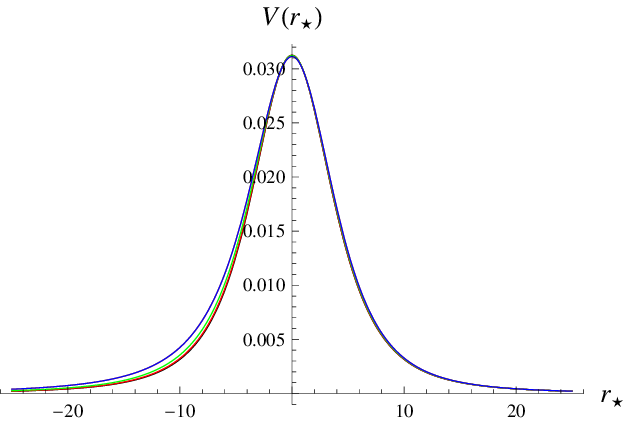}}
\caption{Potential as a function of the tortoise coordinate of the $\ell=0$ scalar field for the Bronnikov extreme black hole with the Minkowski core ($M=1$): $a=0$ (black), $a=0.1$ (red), $a=0.2$ (green), $a=0.3$ (blue).}\label{fig:potentials}
\end{figure}

\begin{figure}
\resizebox{\linewidth}{!}{\includegraphics{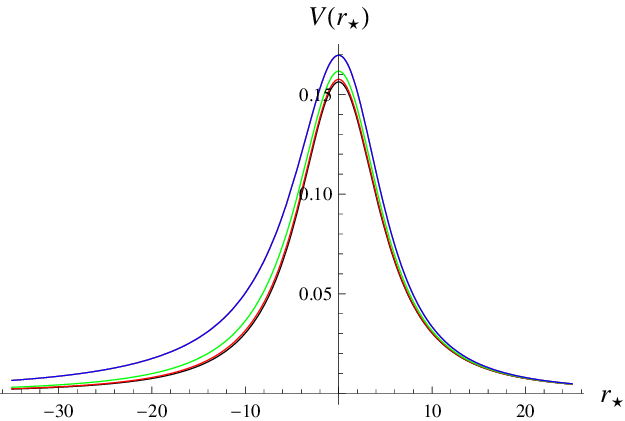}}
\caption{Potential as a function of the tortoise coordinate of the $\ell=1$ scalar field for the Bronnikov extreme black hole with the Minkowski core ($M=1$): $a=0$ (black), $a=0.1$ (red), $a=0.2$ (green),  $a=0.3$ (blue).}\label{fig:potentials2}
\end{figure}

\begin{figure}
\resizebox{\linewidth}{!}{\includegraphics{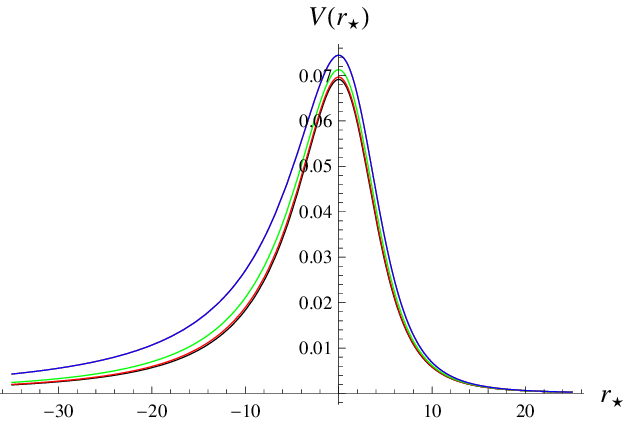}}
\caption{Potential as a function of the tortoise coordinate of the $\ell=1/2$ Dirac field for the Bronnikov extreme black hole with the Minkowski core ($M=1$): $a=0$ (black), $a=0.1$ (red), $a=0.2$ (green),  $a=0.3$ (blue).}\label{fig:potentials3}
\end{figure}

Effective potentials for various values of the parameter $a$ are shown in Figs. \ref{fig:potentials}-\ref{fig:potentials3}. One of the two isospectral effective potentials for the Dirac field has a negative gap near the event horizon, which, due to isospectrality, does not affect the stability properties.

\begin{table*}
\begin{tabular}{c c c c c}
\hline
$a$ & WKB6 Padé & WKB6 &  difference $Re (\omega)$ & difference $Im (\omega)$ \\
\hline
$0$ & $0.237907-0.087749 i$ & $0.237601-0.087395 i$ & $0.129\%$ & $0.403\%$\\
$0.1$ & $0.239301-0.087037 i$ & $0.239027-0.086662 i$ & $0.114\%$ & $0.431\%$\\
$0.2$ & $0.243677-0.084550 i$ & $0.243537-0.084130 i$ & $0.0574\%$ & $0.497\%$\\
$0.3$ & $0.252857-0.078077 i$ & $0.251821-0.078201 i$ & $0.410\%$ & $0.159\%$\\
$0.31$ & $0.253610-0.077410 i$ & $0.252877-0.077258 i$ & $0.289\%$ & $0.197\%$\\
\hline
\end{tabular}
\caption{Quasinormal modes of the $\ell=1/2$, $n=0$ Dirac field for the Bronnikov extreme black hole with the Minkowski core calculated using the 6th WKB method with and without Padé approximants; $M=1$.}
\end{table*}

\begin{table*}
\begin{tabular}{c c c c c}
\hline
$a $ & Prony fit & WKB6 Padé &  rel. error $Re (\omega)$ & rel. error $Im (\omega)$  \\
\hline
$0$ & $0.133319-0.096030 i$ & $0.134025-0.095165 i$ & $0.530\%$ & $0.901\%$\\
$0.05$ & $0.133422-0.095920 i$ & $0.134125-0.095079 i$ & $0.527\%$ & $0.876\%$\\
$0.1$ & $0.133727-0.095581 i$ & $0.134419-0.094814 i$ & $0.518\%$ & $0.802\%$\\
$0.15$ & $0.134221-0.094987 i$ & $0.134888-0.094347 i$ & $0.497\%$ & $0.674\%$\\
$0.2$ & $0.134874-0.094094 i$ & $0.135486-0.093625 i$ & $0.454\%$ & $0.499\%$\\
$0.25$ & $0.135609-0.092845 i$ & $0.136131-0.092544 i$ & $0.385\%$ & $0.323\%$\\
$0.3$ & $0.136270-0.091219 i$ & $0.136690-0.091031 i$ & $0.308\%$ & $0.206\%$\\
$0.31$ & $0.136375-0.090859 i$ & $0.136775-0.090686 i$ & $0.294\%$ & $0.191\%$\\
\hline
\end{tabular}
\caption{Comparison of the quasinormal frequencies for the regular black hole with the Minkowski core obtained by the time-domain integration and the 6th order WKB approach with Padé approximants for $s=\ell=0$ ($M=1$).}\label{check1}
\end{table*}

\begin{table*}
\begin{tabular}{c c c c c}
\hline
$a $ & Prony fit & WKB6 Padé & rel. error $Re (\omega)$ & rel. error $Im (\omega)$  \\
\hline
$0$ & $0.377649-0.089333 i$ & $0.377642-0.089382 i$ & $0.00184\%$ & $0.0556\%$\\
$0.05$ & $0.378109-0.089175 i$ & $0.378102-0.089225 i$ & $0.00178\%$ & $0.0558\%$\\
$0.1$ & $0.379508-0.088687 i$ & $0.379501-0.088737 i$ & $0.00160\%$ & $0.0562\%$\\
$0.15$ & $0.381904-0.087814 i$ & $0.381899-0.087864 i$ & $0.00120\%$ & $0.0565\%$\\
$0.2$ & $0.385405-0.086449 i$ & $0.385404-0.086497 i$ & $0.00020\%$ & $0.0555\%$\\
$0.25$ & $0.390176-0.084390 i$ & $0.390184-0.084445 i$ & $0.00219\%$ & $0.0643\%$\\
$0.3$ & $0.396451-0.081252 i$ & $0.396459-0.081303 i$ & $0.00208\%$ & $0.0620\%$\\
$0.31$ & $0.397909-0.080437 i$ & $0.397919-0.080486 i$ & $0.00249\%$ & $0.0608\%$\\
\hline
\end{tabular}
\caption{Comparison of the quasinormal frequencies for the regular black hole with the Minkowski core obtained by the time-domain integration and the 6th order WKB approach with Padé approximants for $s=0$, $\ell=1$ ($M=1$).}\label{check2}
\end{table*}

\begin{table*}
\begin{tabular}{c c c c c}
\hline
$a $ & Prony fit & WKB6 Padé & rel. error $Re (\omega)$ & rel. error $Im (\omega)$ \\
\hline
$0$ & $0.238146-0.087506 i$ & $0.237907-0.087749 i$ & $0.100\%$ & $0.277\%$\\
$0.05$ & $0.238493-0.087333 i$ & $0.238252-0.087575 i$ & $0.101\%$ & $0.278\%$\\
$0.1$ & $0.239548-0.086792 i$ & $0.239301-0.087037 i$ & $0.103\%$ & $0.282\%$\\
$0.15$ & $0.241351-0.085822 i$ & $0.241091-0.086070 i$ & $0.108\%$ & $0.289\%$\\
$0.2$ & $0.243969-0.084294 i$ & $0.243677-0.084550 i$ & $0.119\%$ & $0.304\%$\\
$0.25$ & $0.247496-0.081969 i$ & $0.247030-0.082236 i$ & $0.188\%$ & $0.326\%$\\
$0.3$ & $0.252017-0.078390 i$ & $0.252857-0.078077 i$ & $0.333\%$ & $0.399\%$\\
$0.31$ & $0.253038-0.077458 i$ & $0.253610-0.077410 i$ & $0.226\%$ & $0.0611\%$\\
\hline
\end{tabular}
\caption{Comparison of the quasinormal frequencies for the regular black hole with the Minkowski core obtained by the time-domain integration and the 6th order WKB approach with Padé approximants for $s=1/2$, $\ell=1/2$ ($M=1$).}\label{check3}
\end{table*}

\begin{figure*}
\resizebox{\linewidth}{!}{\includegraphics{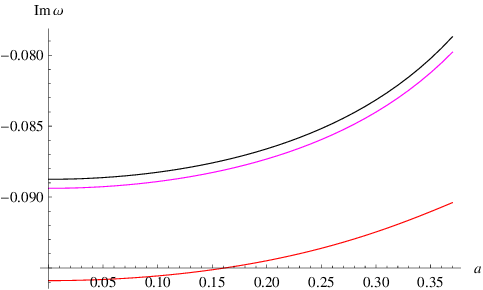}\includegraphics{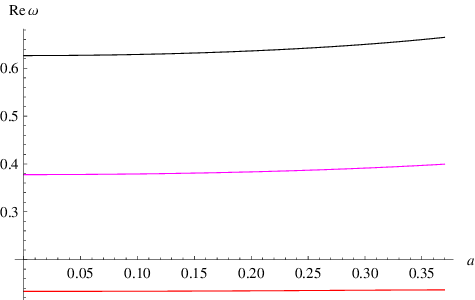}}
\caption{Real and imaginary parts of $\omega$ for $\ell=0, 1, 2$ scalar perturbations for the regular black hole with de Sitter core. Computations are done with the 13th order WKB method with $\tilde{m}=7$.}\label{fig:K2scalar}
\end{figure*}

\begin{figure*}
\resizebox{\linewidth}{!}{\includegraphics{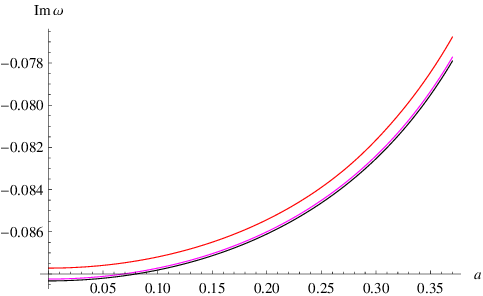}\includegraphics{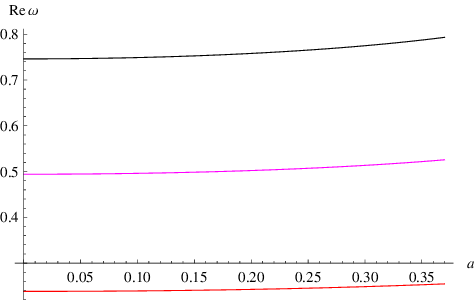}}
\caption{Real and imaginary parts of $\omega$ for $\ell=1/2, 3/2, 5/12$ Dirac perturbations for the regular black hole with de Sitter core. Computations are done with the 13th order WKB method with $\tilde{m}=7$.}\label{fig:K2Dirac}
\end{figure*}

\begin{figure*}
\resizebox{\linewidth}{!}{\includegraphics{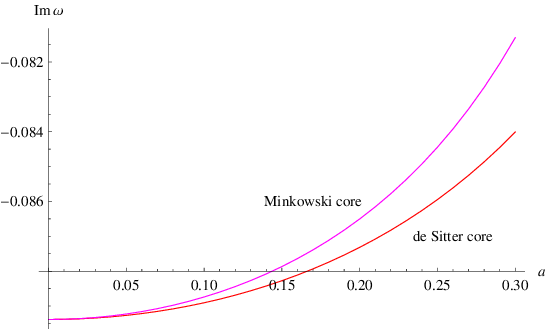}\includegraphics{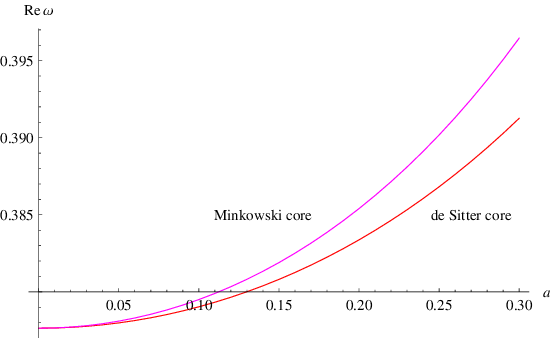}}
\caption{Real and imaginary parts of $\omega$ for $\ell=1$ scalar perturbations for the regular black hole with de Sitter and Minkowski cores. Computations are done with the 13th order WKB method with $\tilde{m}=7$.}\label{fig:L1}
\end{figure*}
\section{WKB method}\label{sec:WKB}

If the effective potential $V(r)$ in the wave equation (\ref{wave-equation}) forms a barrier with a single peak, the WKB method is effective for determining the dominant quasinormal modes, which must satisfy the boundary conditions:
\begin{equation}\label{boundaryconditions}
\Psi(r_*\to\pm\infty)\propto e^{\pm\imo \omega r_*},
\end{equation}
indicating purely ingoing waves at the horizon ($r_*\to-\infty$) and purely outgoing waves at spatial infinity ($r_*\to\infty$).

The WKB method involves matching asymptotic solutions, which satisfy the quasinormal boundary conditions (\ref{boundaryconditions}), with a Taylor expansion around the peak of the potential barrier. The first-order WKB formula corresponds to the eikonal approximation and is exact in the limit $\ell \to \infty$. The general WKB expression for the frequencies is an expansion around the eikonal limit as follows \cite{Konoplya:2019hlu}:
\begin{eqnarray}\label{WKBformula-spherical}
\omega^2&=&V_0+A_2(\K^2)+A_4(\K^2)+A_6(\K^2)+\ldots\\\nonumber
&-&\imo \K\sqrt{-2V_2}\left(1+A_3(\K^2)+A_5(\K^2)+A_7(\K^2)\ldots\right),
\end{eqnarray}
with the quasinormal mode matching conditions given by
\begin{equation}
\K=n+\frac{1}{2}, \quad n=0,1,2,\ldots,
\end{equation}
where $n$ is the overtone number, $V_0$ is the maximum value of the effective potential, $V_2$ is the second derivative of the potential at this maximum with respect to the tortoise coordinate, and $A_i$ (for $i=2, 3, 4, \ldots$) are the WKB correction terms beyond the eikonal approximation, dependent on $\K$ and higher-order derivatives of the potential at its maximum up to order $2i$. The explicit forms of $A_i$ can be found in \cite{Iyer:1986np} for the second and third WKB orders, in \cite{Konoplya:2003ii} for the fourth to sixth orders, and in \cite{Matyjasek:2017psv} for the seventh to thirteenth orders. This WKB approach for determining quasinormal modes has been extensively applied at various orders in numerous studies \cite{Konoplya:2006ar,Kokkotas:2010zd,Balart:2023odm,DuttaRoy:2022ytr,Kodama:2009bf,Al-Badawi:2023lvx,Gonzalez:2022ote,Chen:2023akf}, including the authors' recent analysis \cite{Skvortsova:2023zmj,Skvortsova:2024atk} of quasinormal modes of the $2+1$ dimensional \cite{Konoplya:2020ibi} and quantum corrected \cite{Lewandowski:2022zce} black holes.

\section{Time-domain integration}\label{sec:TD}

The accuracy of the aforementioned WKB method can be validated  by comparing them with an independent approach using time-domain integration. For time-domain integration, we employed the Gundlach-Price-Pullin discretization scheme \cite{Gundlach:1993tp}:
\begin{eqnarray}
\Psi\left(N\right)&=&\Psi\left(W\right)+\Psi\left(E\right)-\Psi\left(S\right)\nonumber\\
&&- \Delta^2V\left(S\right)\frac{\Psi\left(W\right)+\Psi\left(E\right)}{4}+{\cal O}\left(\Delta^4\right),\label{Discretization}
\end{eqnarray}
where the integration points are defined as follows: $N\equiv\left(u+\Delta,v+\Delta\right)$, $W\equiv\left(u+\Delta,v\right)$, $E\equiv\left(u,v+\Delta\right)$, and $S\equiv\left(u,v\right)$. This method has been extensively used in numerous studies \cite{Konoplya:2014lha,Konoplya:2020jgt,Ishihara:2008re,Qian:2022kaq,Varghese:2011ku,Momennia:2022tug} and has proven to be accurate.

To extract the values of frequencies from the time-domain profile, we use the Prony method, which fits the profile data with a sum of damped exponents:
\begin{equation}\label{damping-exponents}
\Psi(t)\simeq\sum_{i=1}^pC_ie^{-i\omega_i t}.
\end{equation}
We assume that the quasinormal ringing stage begins at some initial time $t_0=0$ and ends at $t=Nh$, where $N\geq2p-1$. The relation (\ref{damping-exponents}) is then satisfied for each point of the profile:
\begin{equation}
x_n\equiv\Psi(nh)=\sum_{j=1}^pC_je^{-i\omega_j nh}=\sum_{j=1}^pC_jz_j^n.
\end{equation}
We determine $z_i$ from the known values of $x_n$ and then calculate the quasinormal frequencies $\omega_i$. Quasinormal modes are typically extracted from time-domain profiles when the ringdown stage includes a sufficient number of oscillations. The higher the multipole number $\ell$, the longer the ringdown period.

\begin{figure}
\resizebox{\linewidth}{!}{\includegraphics{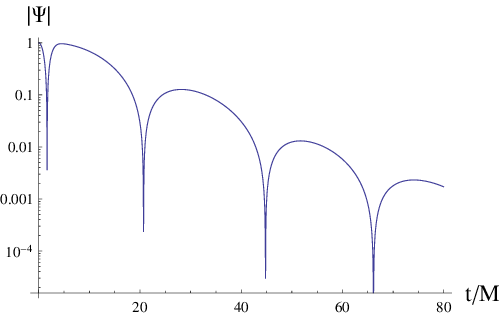}}
\caption{Time-domain profile for the scalar perturbations ($\ell=0$)  hole $a = 0.5$, $M =1$.}\label{fig:timedomain}
\end{figure}

\begin{figure}
\resizebox{\linewidth}{!}{\includegraphics{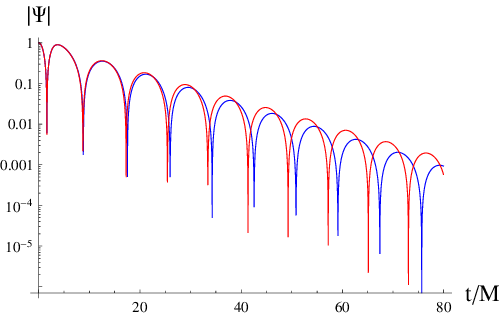}}
\caption{Time-domain profile for the scalar perturbations ($\ell=1$) $a =0.1 $ (blue) and $a =0.3$ (red); $M =1$.}\label{fig:timedomain2}
\end{figure}

\section{Quasinormal modes}\label{sec:QNM}

The quasinormal modes obtained using the 6th order WKB method with Padé approximants and time-domain integration are presented in Tables II-IV for the regular black hole with a Minkowski core. As the parameter $a$ increases, the oscillation frequency shows only a slight increase, while the damping rate significantly decreases. Consequently, the quality factor $Q$,
\begin{equation}
Q \sim \frac{\Re (\omega)}{\Im (\omega)},
\end{equation}
is notably higher for the regular extreme black hole compared to its singular counterpart.

The difference between the results from the Prony fit of the time-domain profiles and the 6th order WKB method with Padé approximants is minimal, usually within a small fraction of one percent. For $\ell > 0$, the standard 6th order WKB method also provides a good approximation, as demonstrated in Table I. However, for $\ell=0$ scalar perturbations, the standard WKB method, even at higher orders, lacks sufficient accuracy and cannot be reliably used. Nonetheless, applying Padé approximants improves the accuracy of the method even for $\ell=0$, as shown in Table II.

Quasinormal modes obtained using the 6th order WKB method with Padé approximants for regular black holes with a de Sitter core are illustrated in Figs. \ref{fig:K2scalar} and \ref{fig:K2Dirac}. We computed frequencies using the 13th order WKB method with $\tilde{m}=7$, as higher order WKB methods show much better agreement with time-domain integration in this case for the near-extreme values of $a$. The time-domain integration yields very close data that cannot be distinguished in the aforementioned plots. Similar to the model with the Minkowski core, the damping rate decreases while the oscillation frequency slightly increases with the introduction of the parameter $a$.

From fig. \ref{fig:L1} one can see that the two models of regular black holes can be always distinguished via their quasinormal spectrum: the model with Minkowski core is characterised by the higher real oscillation frequency and quicker damping rate.

\section{Eikonal formula}\label{sec:Eikonal}

While quasinormal modes are typically calculated numerically, they can be computed analytically in the limit of large multipole numbers $\ell \gg n$, a method employed for various black hole spacetimes \cite{Abdalla:2005hu,Konoplya:2001ji,Paul:2023eep,Bolokhov:2022rqv,Konoplya:2005sy,Dolan:2010wr,Hod:2009td,Jusufi:2020dhz,Breton:2016mqh,Zhidenko:2008fp}. The eikonal regime is intriguing in its own right, as it may introduce new aspects such as eikonal instability \cite{Dotti:2004sh,Takahashi:2011du,Konoplya:2017lhs} or the correspondence between null geodesics and quasinormal modes \cite{Cardoso:2008bp}, which will be discussed later.

Perturbations in a spherically symmetric background can be reduced to a wave-like equation with an effective potential, which can be approximated as follows:
\begin{equation}\label{potential-multipole}
V(r_*)=\kappa^2\left(H(r_*)+\mathcal{O}(\kappa^{-1})\right).
\end{equation}
Here, $\kappa\equiv\ell+\frac{1}{2}$ and $\ell=s,s+1,s+2,\ldots$ is the positive half-integer multipole number, with its minimum value equal to the spin of the field under consideration $s$. Following \cite{Konoplya:2023moy}, we use an expansion in terms of $\kappa^{-1}$.

The function $H(r_*)$ has a single peak, allowing the location of the potential's maximum (\ref{potential-multipole}) to be expanded as:
\begin{equation}\label{rmax}
  r_{\max }=r_0+r_1\kappa^{-1}+r_2\kappa^{-2}+\ldots.
\end{equation}

Substituting (\ref{rmax}) into the first-order WKB formula
\begin{eqnarray}
\omega&=&\sqrt{V_0-\imo \kappa\sqrt{-2V_2}},
\end{eqnarray}
and then expanding in $\kappa^{-1}$, we find:
\begin{eqnarray}\label{eikonal-formulas}
\omega=\Omega\kappa-\imo\lambda\kappa+\mathcal{O}(\kappa^{-1}).
\end{eqnarray}
The above relation provides a reasonable approximation for $\kappa\gg1$.

\begin{widetext}
Expanding in powers of small $a$ we obtain the position of the potential's peak for the solution with the Minkowski core:
\begin{equation}
 r_{\max }=\left(3 M-\frac{5 a }{162 M^3}+O\left(a
   ^2\right)\right)+\frac{O\left(a ^2\right)}{\kappa
   }+O\left(\frac{1}{\kappa }\right)^2.
    \end{equation}
Then, using the first order WKB formula and expansion in powers of $\kappa^{-1}$, we obtain the analytic expression for the quasinormal frequencies in the eikonal regime
 \begin{equation}
\omega = \kappa  \left(\frac{1}{3 \sqrt{3} M}+\frac{5 a }{4374
   \sqrt{3} M^5}+O\left(a
   ^2\right)\right)+\left(-\frac{4 i \K}{9 \sqrt{3}
   M}+\frac{70 i a  \K}{6561 \sqrt{3} M^5}+O\left(a
   ^2\right)\right)+O\left(\frac{1}{\kappa
   }\right),
    \end{equation}
In the way for the solution with the de Sitter core we obtain the position of the potential's maximum:
\begin{equation}
 r_{\max }=\left(3 M-\frac{5 a }{162 M^3}+O\left(a
   ^2\right)\right)+\frac{O\left(a ^2\right)}{\kappa
   }+O\left(\frac{1}{\kappa }\right)^2,
    \end{equation}
and the expression for the quasinormal frequency
 \begin{equation}
 \omega =\kappa  \left(\frac{1}{3 \sqrt{3} M}+\frac{5 a }{4374
   \sqrt{3} M^5}+O\left(a
   ^2\right)\right)+\left(-\frac{4 i \K}{9 \sqrt{3}
   M}+\frac{70 i a  \K}{6561 \sqrt{3} M^5}+O\left(a
   ^2\right)\right)+O\left(\frac{1}{\kappa
   }\right),
    \end{equation}   
\end{widetext}

In \cite{Cardoso:2008bp}, it was demonstrated that the parameters characterizing unstable circular null geodesics around a static and spherically symmetric black hole are dual to the quasinormal modes emitted by the black hole in the regime where $\ell \gg n$: the real and imaginary components of the quasinormal frequency for $\ell \gg n$ are proportional to the frequency and instability timescale of the circular null geodesics, given by:
\begin{equation}\label{QNM}
\omega_n=\Omega\ell-\imo(n+1/2)|\lambda|, \quad \ell \gg n.
\end{equation}
Here, $\Omega$ represents the angular velocity at the unstable null geodesics, and $\lambda$ stands for the Lyapunov exponent.
While this correspondence holds true for many cases, \cite{Konoplya:2017wot} revealed its breakdown when the typical centrifugal term $f(r) \ell (\ell +1)/r^2$ in the effective potential takes on a different form, as seen, for instance, in Einstein-Gauss-Bonnet theories, which generally permit a scalar field \cite{Konoplya:2017wot,Konoplya:2020bxa,Konoplya:2019hml}. Although it was initially shown that the correspondence basically holds as long as the WKB formula by \cite{Schutz:1985km} for quasinormal modes remains valid \cite{Konoplya:2017wot}, it was later discovered that even in such cases, the full eikonal spectrum might not be accurately reproduced by the WKB formula, consequently leading to discrepancies in the overtone number $n$ and the real frequency \cite{Konoplya:2022gjp,Bolokhov:2023dxq}.

Here, one can observe that when null geodesics are utilized, the aforementioned correspondence is fulfilled. However, in the context of non-linear electrodynamics, photons do follow the null geodesics, as noted in \cite{Toshmatov:2019gxg}. Thus, while the correspondence formally holds in this scenario, it's important to remember that "null" no longer implies photons.

\section{Conclusions}\label{sec:conclusions}

In this study, we examined the quasinormal modes of massless scalar and Dirac perturbations around regular, maximally charged black holes recently discovered within the framework of nonlinear modifications of Maxwell electrodynamics \cite{Bronnikov:2024izh}. Calculations were performed using two independent methods: the higher order WKB method and time-domain integration, yielding a good agreement between them.

The results from time-domain integration indicate stability against linear perturbations for the fields, with a better quality factor observed for these black holes compared to their counterparts in linear electrodynamics. In the eikonal limit, we derived analytical expressions for quasinormal modes and discussed the limits of correspondence between null geodesics and eikonal quasinormal frequencies.

An interesting question, which was beyond the scope of our study, is the investigation of coupled perturbations, such as perturbations of electromagnetic or gravitational fields. This problem requires a special study on the reduction of perturbation equations to a wave-like form and merits separate consideration.

\begin{acknowledgments}
I would like to acknowledge R. A. Konoplya and S. V. Bolokhov for useful discussions. This work was supported by RUDN University research project FSSF-2023-0003.
\end{acknowledgments}

\bibliographystyle{unsrt}
\bibliography{bibliography}
\end{document}